\documentclass{article}
\usepackage{setspace}
\usepackage{spconf,graphicx}
\usepackage{amsmath}
\usepackage{amsfonts}

\title{AUDIO CODING WITH UNIFIED NOISE SHAPING AND PHASE CONTRAST CONTROL}
\name{Byeongho Jo, Seungkwon Beack, and Taejin Lee
\thanks{This work was supported by Electronics and Telecommunications Research Institute (ETRI) grant funded by the Korean government [22ZH1200, The research of the basic media - contents technologies].}}
\address{Electronics and Telecommunications Research Institute, Republic of Korea \\ 
Email: \{bhjo, skbeack, tjlee\}@etri.re.kr}
\begin{document}
\maketitle
\begin{abstract}
Over the past decade, audio coding technology has seen standardization and the development of many frameworks incorporated with linear predictive coding (LPC). As LPC reduces information in the frequency domain, LP-based frequency-domain noise-shaping (FDNS) was previously proposed. To code transient signals effectively, FDNS with temporal noise shaping (TNS) has emerged. However, these mainly operated in the modified discrete cosine transform domain, which essentially accompanies time domain aliasing. In this paper, a unified noise-shaping (UNS) framework including FDNS and complex LPC-based TNS (CTNS) in the DFT domain is proposed to overcome the aliasing issues. Additionally, a modified polar quantizer with phase contrast control is proposed, which saves phase bits depending on the frequency envelope information. The core coding feasibility at low bit rates is verified through various objective metrics and subjective listening evaluations. 
\end{abstract}

%%%%%%%%%%%%%%%%%%%%%%%%%%%%%%%%%%%%%%%%%%%%%%%%%%%%%%%%%%%%%%%%%%%%%%%%%%%%%%%%%%%%%%
%%%%%%%%%%%%%%%%%%%%%%%%%%%%%%%%%%%%%%%%%%%%%%%%%%%%%%%%%%%%%%%%%%%%%%%%%%%%%%%%%%%%%%
\section{Introduction}
Psychoacoustic-based audio coding has evolved as a method to preserve perceptual quality while decreasing bit rates \cite{bosi2002introduction}. In contrast, speech coding has been refined using linear prediction (LP) models \cite{gibson2005speech,AMRWBp}. Separately developed audio and speech coding evolved into the unified speech and audio coding frameworks (MPEG USAC \cite{usac}), such as LP-based frequency-domain noise shaping (FDNS). In addition, LP models are used to reduce temporal information by temporal noise shaping (TNS), which is a representative tool conforming to the audio coding standards \cite{herre1996enhancing,AAC}. The sequential combination of FDNS and TNS in the modified discrete cosine transform (MDCT) domain \cite{fuchs2015low} is feasibly realized in enhanced voice services (EVS) \cite{evs} and MPEG-H 3D audio \cite{mpegh}.

TNS is mainly incorporated in MDCT, which may incur the undesired time-domain aliasing noises after decoding \cite{herre1999temporal,liu2008compression}. Specifically, the high noise level at the edge of an attack signal, time-domain aliasing noise, and noise spread due to the order of the TNS filter are three undesired artifacts, which can be reduced by window-switching or using a small overlapped window \cite{allamanche1999mpeg}. However, the window-switching increases the complexity and algorithmic delay. In addition, using a small overlapped window can minimize the aliasing noise but does not avoid it fundamentally \cite{herre1996enhancing}. Thus, this study presents a unified noise shaping (UNS) model in the time and frequency domain without any aliasing-related artifacts.

In the UNS framework, FDNS and TNS are sequentially operated using discrete Fourier transform (DFT). Specifically, a complex linear predictive coding (LPC)-based TNS (CTNS) model is operated, which predicts a temporal envelope close to the Hilbert envelope. The model effectively reduces the amount of temporal information, but the residual signal has a complex value. Hence, an efficient quantization of the complex value should be developed. Exclusively, the UNS framework with the DFT can reduce the structural complexity, as it does not perform any additional inverse DFT or MDCT in the encoder. Since the quantization target of the UNS framework is a complex-valued signal, several quantizers in the polar coordinates should be considered, such as a strict polar quantizer (SPQ) \cite{pearlman1979polar}, a conventional polar quantizer \cite{neuhoff1997polar}, and an unrestricted polar quantizer (UPQ) \cite{wilson1980magnitude}. Among them, the UPQ utilizes magnitude levels for phase quantization, and its performance outperforms the other polar quantizers \cite{vafin2005entropy}. Recently, an entropy-constrained UPQ (ECUPQ) and its optimal solution at low rates were reported \cite{vafin2005entropy,wu2018design}. Yoon and Malvar presented an audio coding framework with a UPQ for modulated complex lapped transform (MCLT) coefficients \cite{yoon2008coding}. The authors'  previous study proposed a modified UPQ (mUPQ) combined with FDNS at low bit rates \cite{Jo2022modified}. Thus, this study presents an extended mUPQ with a phase contrast control (PCC) utilizing the frequency envelope ratio (FER) calculated from the quantized LPC model.

\label{sec:intro}

\section{TNS with MDCT}
TNS is used to avoid pre-echo artifacts, which may occur when coding a transient signal with a long window \cite{herre1996enhancing}. Considering the duality between the spectral and Hilbert envelopes, TNS is operated with an open-loop prediction in the frequency domain. TNS can be adapted as an independent tool in any audio coder, and it works effectively for "pitched" signals where the window-switching tool does not operate well \cite{herre1996enhancing}. However, when TNS is operated in the MDCT domain, it may cause the undesired time-domain aliasing noises \cite{liu2008compression}. The MDCT operation can be decomposed into a time-domain aliasing block and shifted DFT \cite{wang2000relationship}. Hence, the MDCT coefficient predictions are accompanied by time-domain aliased terms, such that the estimated temporal envelope is distorted. The artifacts can be avoided fundamentally by using CTNS in the DFT domain. 

\begin{figure}[t]
\begin{minipage}[b]{1.0\linewidth}
  \centering
  \centerline{\includegraphics[width=7.5cm]{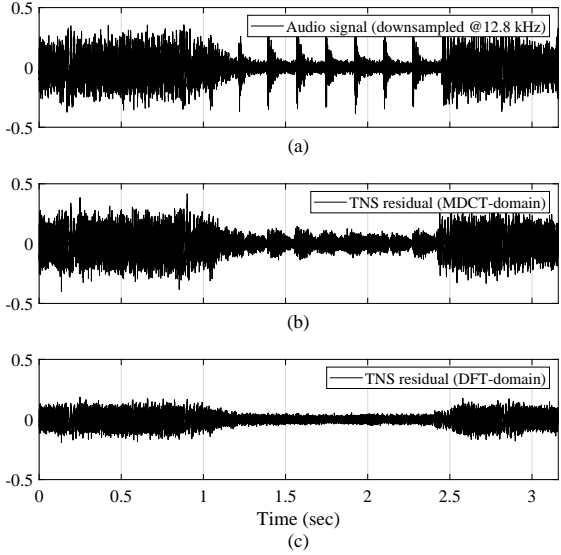}}
%  \vspace{2.0cm}
  % \centerline{(a) Result 1}\medskip
\end{minipage}
\caption{Illustration of TNS in two different domains. The (a) downsampled audio signal (music1), (b) residual signal in the MDCT domain, and (c) residual signal in the DFT domain}
\label{fig:CTNS_effects}
\end{figure}

Figure \ref{fig:CTNS_effects} shows the reconstructed TNS residual signals in the (b) MDCT and (c) DFT domains. For analyzing and synthesizing the TNS residual signals in the two domains, the same window configuration was utilized. The TNS orders were set to $16$. Since the prediction is incomplete in the MDCT domain, the TNS residual signal with the MDCT in the transient period has relatively high power ($1\sim2.5$ sec). Contrarily, the TNS residual signal with the DFT in the transient period is whitened in the time domain, which implies that the Hilbert envelope is well predicted by the CTNS. TNS with DFT reduces the temporal information, which leads to a decrease in the entropy of the quantized coefficients.
\label{sec:limits}

\section{Proposed Audio Coding System}
\label{sec:pro}
\subsection{Overview}
\begin{figure}[t]
	% \begin{minipage}[b]{1.0\linewidth}
	\centering
	% \centerline{
	\includegraphics[width=8.5cm]{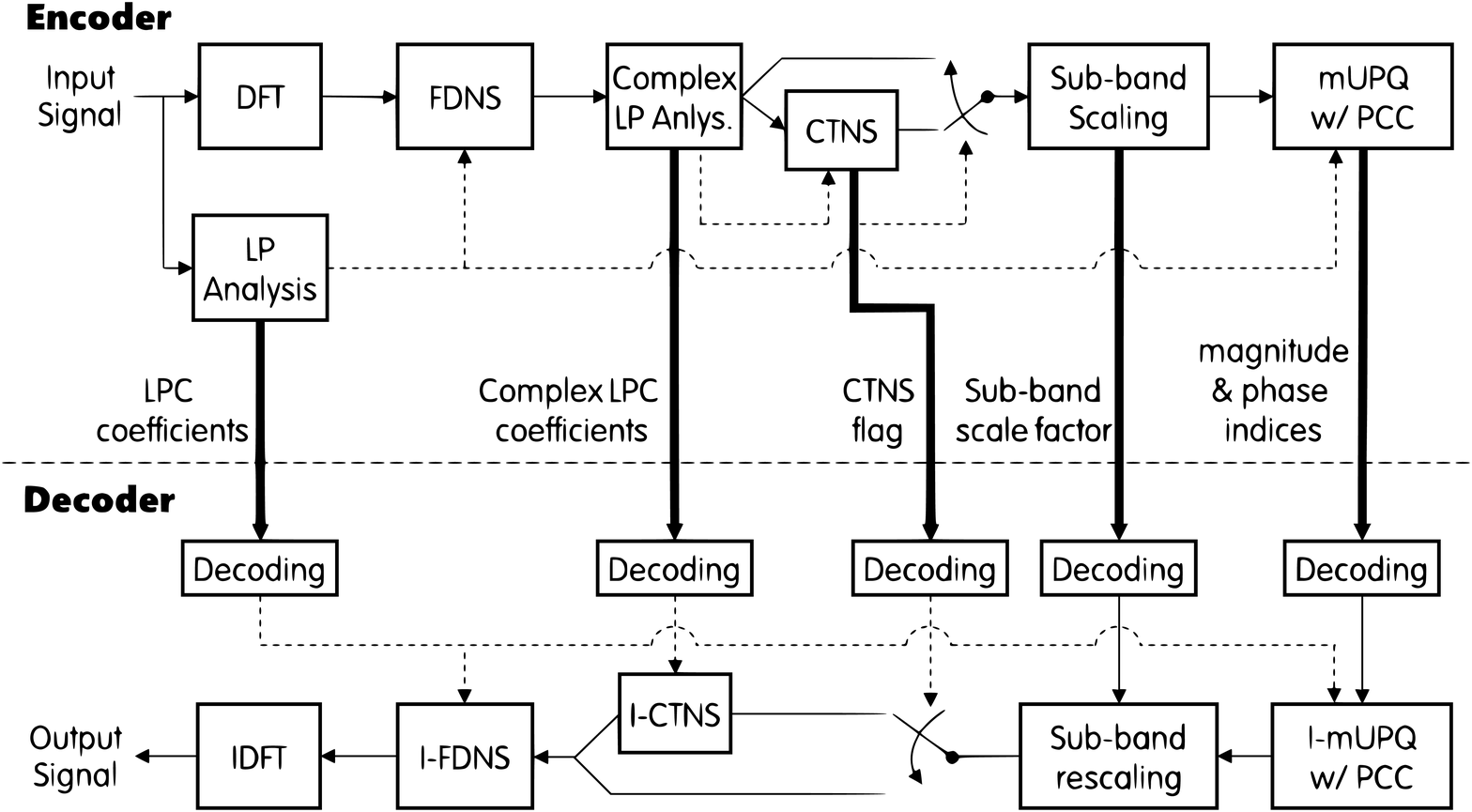}
	%  \vspace{2.0cm}  
	% \end{minipage}
	\caption{Block diagram of the encoder and decoder. Thin and solid lines indicate signal flows. Dotted lines indicate control signals. The encoded signals are described as thick and solid lines.}
	\label{fig:block}
\end{figure}

Figure \ref{fig:block} shows the proposed audio coding system. Unlike the transform coded excitation (TCX) in USAC \cite{fuchs2009mdct}, the residual signal is computed by the UNS using DFT. First, the audio signal is windowed and fed to the DFT and LP analysis blocks simultaneously. For DFT, the window can be chosen without considering the time domain aliasing cancellation (TDAC) condition. Therefore, a small overlap window without zero sides is used. Unlike the FDNS of MPEG USAC \cite{usac}, the FDNS of the proposed system is done by dividing the magnitude of the DFT coefficients by the reconstructed frequency envelope (FE) generated using the quantized LPC model without interpolation. Second, the complex-valued FDNS residual signal undergoes complex LP analysis, which is performed using the same LP algorithm \emph{e.g.} Levinson-Durbin recursion. The filtering process using the complex LPC is conducted above $312$ Hz. The filtered signal ($x_{CT}\in \mathbb{C}$) is utilized for calculating the predicted gain as
\begin{equation}
G(n) = 10\log_{10}{\frac{\sum_f\left| x_{FD}(n,f)- x_{CT}(n,f) \right|^2}{\sum_f\left| x_{FD}(n,f)\right|^2}},
\end{equation}
where $n$ and $f$ are the frame and frequency indices, respectively, and the complex value $x_{FD}$ denotes the FDNS residual signal. The CTNS block operates when the prediction gain is higher than the pre-defined threshold. The flag indicating the on/off state of the CTNS block for each frame is allocated a $1$ bit and transmitted to the decoder. 

\begin{figure}[t]
\begin{minipage}[b]{1.0\linewidth}
  \centering
  \centerline{\includegraphics[width=8.5cm]{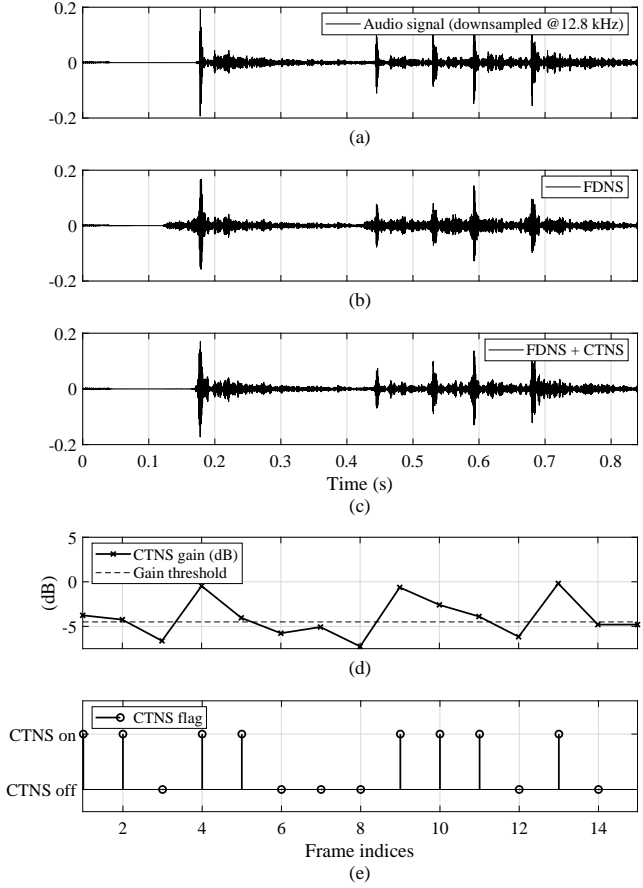}}
%  \vspace{2.0cm}
  % \centerline{(a) Result 1}\medskip
\end{minipage}
\caption{Illustration of CTNS and CTNS determination. The (a) downsampled audio signal (cast01), (b) reconstructed signal with FDNS, (c) reconstructed signal with FDNS and CTNS, (d) plot of CTNS gain per frame indices and its threshold, and (e) plot of CTNS flag per frame indices.}
\label{fig:CTNS_on_off}
\end{figure}

The effects of the proposed UNS are illustrated in Fig. \ref{fig:CTNS_on_off}. For all simulations, the same window configuration was utilized. It is observed that the reconstructed signal with only FDNS contains quantization noises spread around the attack period, shown in Fig. \ref{fig:CTNS_on_off} (b). Conversely, the reconstructed signal with the proposed UNS contains well-shaped quantization noises as shown in Fig. \ref{fig:CTNS_on_off} (c). Figures \ref{fig:CTNS_on_off} (d) and (e) show the prediction gain and CTNS flag for each frame, respectively. In this study, the threshold for predicting gain was set to $-4.5$ dB.

Before quantization, the complex-valued UNS residual signal is divided by a scale factor for each sub-band to reduce the dynamic range perceptually. The boundaries of the sub-bands are set to [$40$, $90$, $140$, $200$, $260$, $330$, $410$, $512$], which is the modified equivalent-rectangular bandwidth (ERB) scale. The scale factor for each sub-band is determined by the allocated bits and power of each sub-band \ttfamily{SQ\_gain}\normalfont{} in MPEG USAC \cite{usac}. It is uniformly quantized in the decibel scale with $1$ dB step, and then entropy coded. The input bits for the \ttfamily{SQ\_gain}\normalfont{} were set to [$45$, $34$, $30$, $23$, $19$, $16$, $16$, $16$] for $12$ kbps and [$67$, $50$, $45$, $34$, $29$, $23$, $23$, $23$] for $16$ kbps. The scaled residual signal is then quantized by mUPQ with PCC. The detailed quantization procedure is elaborated in section \ref{sec:mUPQ}. The quantized magnitude and phase indices are entropy coded and transmitted to the decoder. Since the methodology of memory-efficient entropy coding is not the main focus of this study, the sample entropies of the adjacent four indices within the same frame were computed to coordinate the target bits.

The decoder used is shown at the bottom of Fig. \ref{fig:block}. Decoding procedures are opposite to that of an encoder. The decoded magnitude and phase indices are fed to the inverse mUPQ, which produces a de-quantized signal. The de-quantized complex signal is rescaled for each sub-band by multiplying it with the decoded scale factor. The rescaled signal is optionally fed to the inverse CTNS filter when the decoder CTNS flag is on. The inversely filtered or bypassed signals are multiplied by the reconstructed FE computed from the quantized LPC. Finally, the output signal is reconstructed by using inverse DFT (IDFT).

\subsection{Polar quantizer with PCC}\begin{figure}[t]
\begin{minipage}[b]{1.0\linewidth}
  \centering
  \centerline{\includegraphics[width=8.5cm]{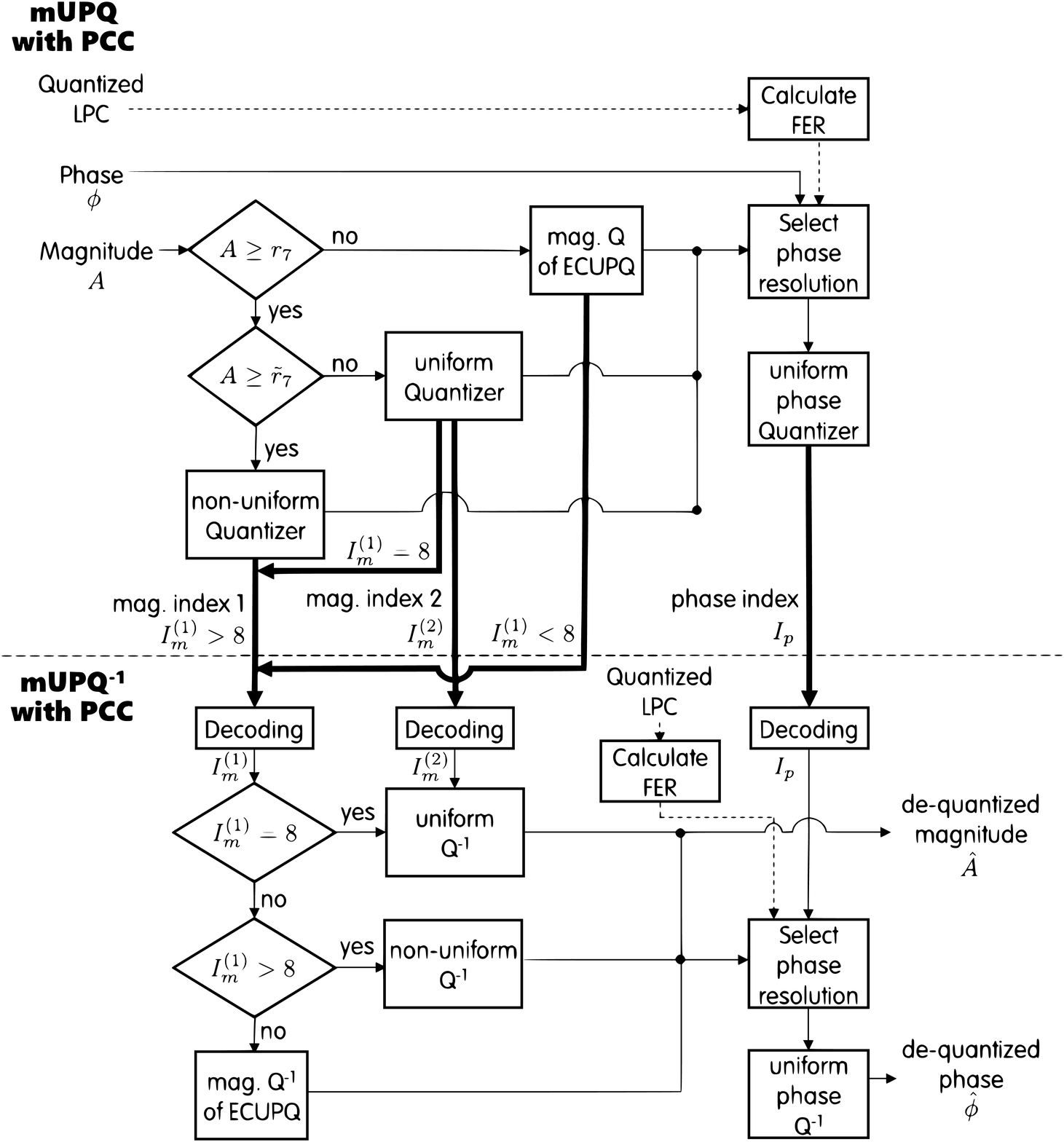}}  
\end{minipage}
\caption{Schematic of mUPQ with PCC and its counterpart (de-quantizer). Thin and solid lines indicate signal flows. Dotted lines indicate control signals. The encoded signals are described as thick and solid lines.}
\label{fig:quant}
\vspace{-0.3cm}
\end{figure}

The magnitude and phase of the complex-valued UNS residual signal are sequentially quantized by mUPQ with PCC. The magnitude is mostly quantized by the ECUPQ solution at a rate of $2.495$ bits/samples reported in \cite{wu2018design}. However, when the ECUPQ is exclusively used, the local spectral peaks may not be precisely reconstructed as presented in \cite{Jo2022modified}. Alternatively, a hybrid quantization is adopted, as shown in Fig. \ref{fig:quant}. The magnitude is quantized by the ECUPQ when its value is less than the highest threshold ($r_7 = 5.056$) of the ECUPQ. Then, the magnitude index $1$ is obtained. Otherwise, the magnitude is determined by comparing its value to be greater than or equal to the second threshold ($\tilde{r}_7= 8.5^{4/3}$). In the former, the magnitude quantization index $1$ ($I^{(1)}_m$) is computed as a nonlinear equation.
\begin{equation}
I^{(1)}_m = {\left\lfloor {{A^{3/4}} + 0.5} \right\rfloor}.
\label{eq:nonSQ}
\end{equation}
While in the latter, the magnitude is uniformly quantized by a rounding operation, where the magnitude quantization index $2$ is calculated ($I^{(2)}_m$). The magnitude indices ($I^{(1)}_m$ and $I^{(2)}_m$) are entropy coded and then transmitted to the decoder. 

The phase is uniformly quantized because the phases are assumed to be a uniformly distributed random variable \cite{pearlman1979polar}. To determine the number of phase cells, the conventional UPQ only considers its magnitude. However, the proposed PCC additionally considers the FER calculated from the quantized LPC to control the phase contrast. The FER for each sub-band is calculated by the reconstructed FE, $H_{dB}(n,f)$ in decibels, as 
\begin{equation}
FER(n,b) = \frac{\bar{H}_{dB}(n,b)}{\sum_{b}{\bar{H}_{dB}(n,b)}},
\end{equation}
where the maximum FE per sub-band is computed as
\begin{equation}
\bar{H}_{dB}(n,b) = \max_{f \in B_b}{H_{dB}(n,f)},
\end{equation}
Here, $b$ denotes the sub-band index and $B_b$ indicates the set of frequency indices within the $b$-th sub-band. For a sub-band with FER greater than the FER threshold, the number of phase-cells set for the corresponding sub-band is assigned as [$1$, $8$, $16$, $16$, $32$, $32$, $64$, $64$] for uniform quantization, else as [$1$, $4$, $8$, $8$, $16$, $16$, $32$, $32$], which has the half resolution. The FER threshold is set to $0.125$, such that the segmental signal-to-noise ratio (SNR) is maximum. Depending on the FER, which contains the coarsely modeled FE information, the phase resolution is controlled by allocating fewer bits to the sub-band with FE relatively small within the frame. This reduces the phase bits for perceptually less dominant spectral components.

The decoder first decodes magnitude index $1$ and then feds it to the hybrid magnitude de-quantizer. When the magnitude index $1$ equals to $8$ indicating that the magnitude is within the outlier region, the magnitude is reconstructed by magnitude index $2$ directly. Else, inverse quantization is performed according to the ECUPQ solution or inversion procedure of nonlinear quantization, depending on the magnitude index $1$. For nonlinear de-quantization, the magnitude is reconstructed to the power of $4/3$. To determine the number of phase cells for uniform phase de-quantizer, the usage of the de-quantized magnitudes and calculated FER is similar to that in the encoder. 

\label{sec:mUPQ}
% inverse quantization side & reconstruction 설명
% FER threshold 어떻게 결정했는지
% reconstruction은 어떻게 하는지
%%%%%%%%%%%%%%%%%%%%%%%%%%%%%%%%%%%%%%%%%%%%%%%%%%%%%%%%%%%%%%%%%%%%%%%%
% DFT + FDNS (o)
% CTNS 작동 및 on/off 기준 (o)
% sub-band scaling 설명 (o)
% entropy coding 어떻게 했는지 설명해야함 (o)
% in the time domain --> in time domain
% 제안 구조가 TCX와 유사하지만, DFT 도메인에서 FDNS --> CTNS 순으로 수행된다는 점이 가장 큰 차이점이라고 언급해주어야 함.
%%%%%%%%%%%%%%%%%%%%%%%%%%%%%%%%%%%%%%%%%%%%%%%%%%%%%%%%%%%%%%%%%%%%%%%%

\section{Performance evaluation}
\subsection{Setup}
% 비트율, 사용한 신호, 인터널 샘플 레이트, DFT 셋업, LPC 계수 오더 및 양자화 방법, 
Objective evaluation and subjective listening tests were conducted to assess the coding performance. For objective evaluation, segmental SNR (segSNR), noise-to-mask ratio (NMR), and perceptual evaluation of audio quality (PEAQ) were calculated \cite{itu1998method,kabal2002examination}, while the listening test was conducted based on the MUSHRA methodology \cite{recommendation2001method}. The tests were conducted with $10$ experienced listeners using reference quality headphones, and $10$ audio signals including $3$ speeches, $4$ music scores, and $3$ mixed items were extracted from the MPEG audio sequences. Bit rates at $12$ kbps and $16$ kbps were selected for evaluation. For comparison, the TCX of MPEG USAC was chosen as the base \cite{usac}. To evaluate the coding performance of the core-band primarily, noise-filling and bandwidth extensions were deactivated. The core-band was coded with a downsampled audio signal at $12.8$ kHz. 

A cosine-tapered window of $1024$ samples with $256$-sized overlap was used for DFT analysis and synthesis. The LPC orders for both FDNS and CTNS were set to $16$, while the perceptual weights for LP analysis were used at $0.98$ and $0.9$, respectively. The LPC for FDNS was transformed to line spectral frequencies (LSFs) and quantized using a two-stage vector quantizer (VQ) with codewords of $20$ bits. Conversely, the complex LPC for CTNS was directly quantized using a three-stage VQ with codewords of $30$ bits for the magnitude and phase, respectively. The allocated bits are shown in Table \ref{tab:bit}. The bits of the complex LPC for CTNS can be efficiently reduced by CTNS switching, similar to the conventional TNS.

\subsection{Evaluation results}\begin{table}[t!]
\caption{Bit allocation for the proposed audio coding system}\medskip
\begin{minipage}[b]{1.0\linewidth}
  \centering
  \centerline{\includegraphics[width=6cm]{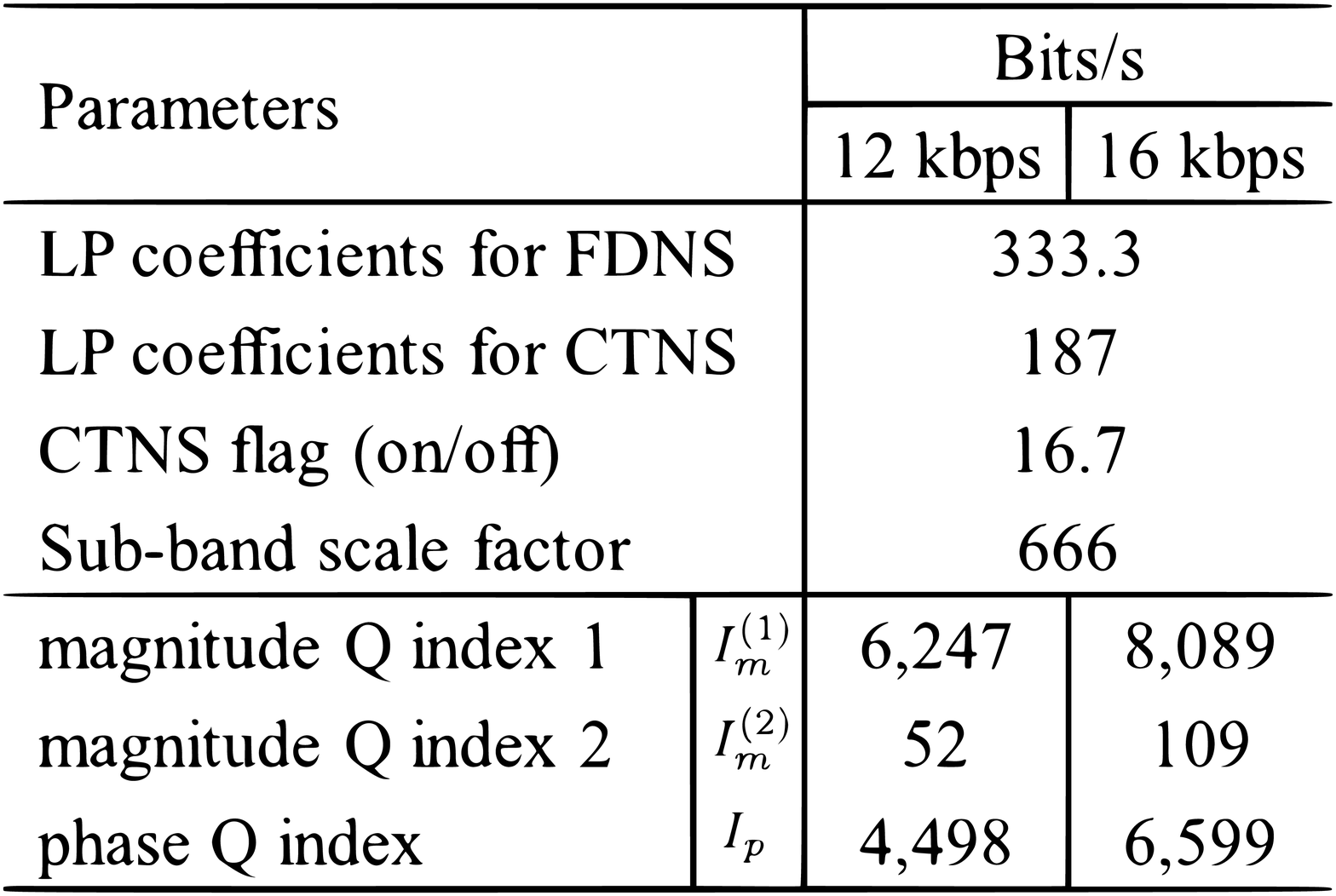}}
 \vspace{-0.65cm}  
\end{minipage}
\label{tab:bit}
\end{table}
\begin{table}[t!]
\caption{Objective scores of TCX with USAC and the proposed system for bit rates of $12$ kbps and $16$ kbps}
\begin{minipage}[b]{1.0\linewidth}  
  \centering
  \centerline{\includegraphics[width=7cm]{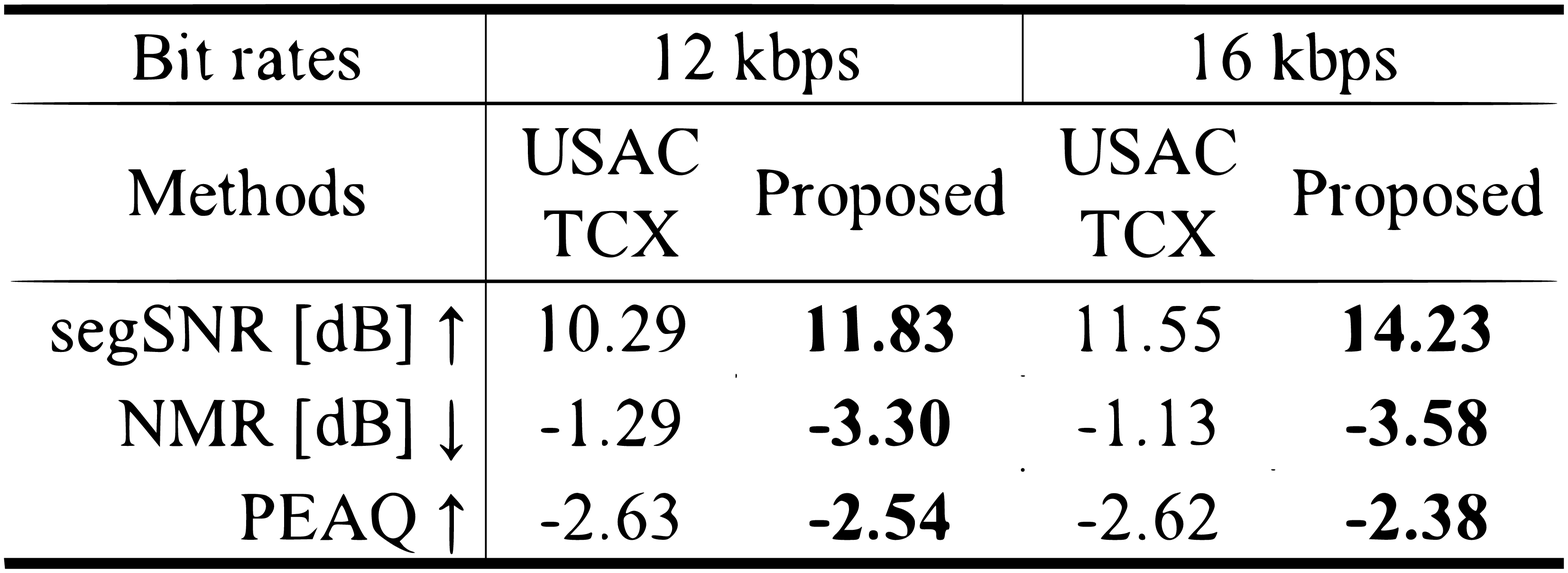}}  
\end{minipage}
\label{tab:objmea}
\end{table}
\begin{figure}[t]
\begin{minipage}[b]{1.0\linewidth}
  \centering
  \centerline{\includegraphics[width=8.5cm]{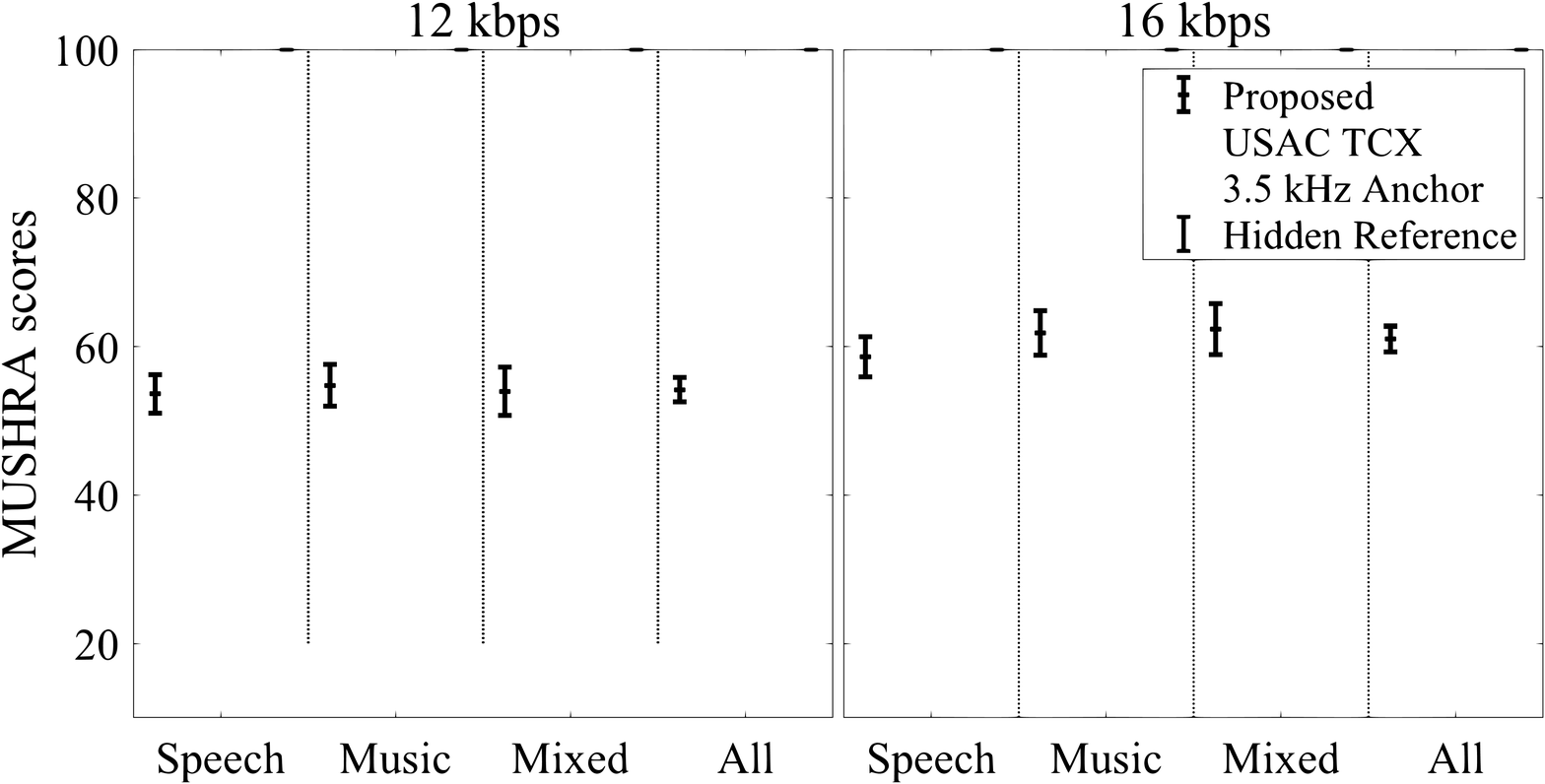}}
\end{minipage}
\caption{MUSHRA scores and their $95$\% confidence intervals at (a) $12$ kbps and (b) $16$ kbps.}
\label{fig:mushra}
 \end{figure}

The objective scores are summarized in Table \ref{tab:objmea}, where the scores of the proposed system outperformed the baseline. The score differences of the three metrics at $16$ kbps were higher than those at $12$ kbps. Figure \ref{fig:mushra} shows the MUSHRA scores of the proposed system and the baseline with a $95$ \% confidence interval. At $12$ kbps, the coding performances of the proposed system for speech and music were statistically better than those of the baseline. For music, though the confidence intervals of both systems overlap, the proposed system mean was higher than the baseline. At $16$ kbps, the coding performance of the proposed system for speech was statistically better than that of the baseline. Nonetheless, the mean of the proposed system for both music and mixed audio was higher than that of the baseline. Thus, the proposed system has statistically better sound quality at $12$ and $16$ kbps for all audio items. Although different window configurations are considered against the baseline, the proposed DFT-based audio coding notably produces a perceptually better sound at low bit rates compared to the conventional MDCT-based coding system.

\label{sec:eval}
\section{Conclusion}
In this paper, an audio coding framework that includes UNS and mUPQ with PCC in the DFT domain is proposed. The proposed UNS model is composed of FDNS without interpolation and CTNS model in the DFT domain, which efficiently reduces the time and frequency-domain information during transient periods without window-switching. The UNS residual signal is quantized by an mUPQ with PCC, which selectively economizes the phase bits based on the computed FER for sub-bands. The objective metric scores of the proposed system outperform those of the baseline. In addition, the total scores of the listening tests are statistically better than those of the baseline at lower bit rates. 
%%%%%%%%%%%%%%%%%%%%%%%%%%%%%%%%%%%%%%%%%%%%%%%%%%%%%%%%%%%%%%%%%%%%%%%%%%%%%%%%%%%%%%
\clearpage
% \linespread{0.75}\selectfont
% \let\oldthebibliography=\thebibliography
% \let\endoldthebibliography=\endthebibliography
% \renewenvironment{thebibliography}[1]{%
%    \begin{oldthebibliography}{#1}%
%      \setlength{\itemsep}{-.4ex}%
% }%
% {%
%    \end{oldthebibliography}%
% }
% \newcommand{\BIBdecl}{\setlength{\itemsep}{-2.0ex}}
\bibliographystyle{ieeebib.bst}
\bibliography{refs.bib}
\end{document}